\documentclass[traditabstract]{aa}

\usepackage{lscape,graphicx,natbib,moreverb,verbatim,fancyvrb,txfonts}

\newcommand{\figdir}
  {./}
\newlength{\figwidth}
\newcommand{\sect}[1]
  {Section~\ref{section:#1}}
\newcommand{\fig}[1]
  {Fig.~\ref{figure:#1}}
\newcommand{\tabl}[1]
  {Table~\ref{table:#1}}

\newcommand{\unit}[1]
  {{\mbox{\rm\,#1}}}
\newcommand{\percent}
  {\,{\rm{per\ cent}}}
\newcommand{\lya}
  {{{\rm{Ly}}\,\mbox{$\alpha$}}}
\newcommand{\lyb}
  {{{\rm{Ly}}\,\mbox{$\beta$}}}
\newcommand{\hii}
  {H\,{\sc{ii}}}
\newcommand{\mgii}
  {Mg\,{\sc{ii}}}

\newcommand{\nv}
  {N\,{\sc{v}}}
\newcommand{\civ}
  {C\,{\sc{iv}}}
\newcommand{\siiv}
  {Si\,{\sc{iv}}}
\newcommand{\ciii}
  {C\,{\sc{iii}}]}

\newcommand{\angstrom}
  {{\mbox{\rm{\AA}}}}
\newcommand{\micron}
  {{\mbox{$\mu${\rm{m}}}}}

\newcommand{\uprime}
  {\mbox{$u$}}
\newcommand{\gprime}
  {\mbox{$g$}}
\newcommand{\rprime}
  {\mbox{$r$}}
\newcommand{\iprime}
  {\mbox{$i$}}
\newcommand{\zprime}
  {\mbox{$z$}}

\newcommand{\zmy}
  {\mbox{\zprime$-$$Y$}}
\newcommand{\ymj}
  {\mbox{$Y$$\!\,-$$J$}}

\newcommand{\imy}
  {\mbox{\iprime$-$$Y$}}

\newcommand{\ston}
  {$S/N$}
\newcommand{\sqdeg}
  {\mbox{${\rm{deg}}^2$}}
\newcommand{\zabs}
  {{z_{\rm{abs}}}}

\newcommand{\fluxobs}
  {f_{\lambda,{\rm{obs}}}}
\newcommand{\fluxintr}
  {f_{\lambda,{\rm{int}}}}
\newcommand{\taueff}
  {\tau_{\rm{eff}}}
\newcommand{\mabs}
  {M_{1450,{\rm{AB}}}}
\newcommand{\ab}
  {{\rm{AB}}}
\newcommand{\vega}
  {{\rm{Vega}}}
\newcommand{\simm}
  {\sim \!}

\newcommand{\firstsdssqso}
  {SDSS~0836$+$0054}
\newcommand{\secondsdssqso}
  {SDSS~1411$+$1217}

\newcommand{\firstqso}
  {ULAS~J0203$+$0012}
\newcommand{\firstqsolong}
  {ULAS~J020332.38$+$001229.2}
\newcommand{\zfirstqsoold}
  {5.86}
\newcommand{\zfirstqso}
  {5.72}
\newcommand{\secondqso}
  {ULAS~J1319$+$0950}
\newcommand{\secondqsolong}
  {ULAS~J131911.29$+$095051.4}
\newcommand{\zsecondqso}
  {6.13}

\newcommand{\etal}
  {et al.}
\newcommand{\ie}
  {i.e.}
\newcommand{\eg}
  {e.g.}

\newcommand{\cf}
  {cf}

\newcommand{\probqso}
  {P_{\rm{q}}}

\begin{document}

\title{Discovery of a  redshift 6.13 quasar in 
  the UKIRT Infrared Deep Sky Survey}
\titlerunning{A redshift 6.13 quasar in UKIDSS}

\author{
  D.\ J.\ Mortlock \inst{1} \and
  M.\ Patel \inst{1} \and
  S.\ J.\ Warren \inst{1} \and 
  B.\ P.\ Venemans \inst{2} \and
  R.\ G.\ McMahon \inst{2} \and
  P.\ C.\ Hewett \inst{2} \and
  C.\ Simpson \inst{3} \and
  R.\ G.\ Sharp \inst{4} \and
  B.\ Burningham \inst{5} \and 
  S.\ Dye \inst{6} \and
  S.\ Ellis \inst{7} \and
  E.\ A.\ Gonzales--Solares \inst{2} \and
  N.\ Hu\'elamo \inst{8} 
}
\authorrunning{D.\ J.\ Mortlock \etal}

\institute{
  Astrophysics Group, Imperial College London, Blackett Laboratory, 
  Prince Consort Road, London, SW7 2AZ, UK\\ 
  \email{mortlock@ic.ac.uk}
\and
  Institute of Astronomy, University of Cambridge, 
  Madingley Road, Cambridge, CB3 0HA, UK
\and
  Astrophysics Research Institute, Liverpool John Moores
  University, Twelve Quays House, Egerton Wharf, Birkenhead, CH41 1LD, UK
\and
  Anglo Australian Observatory, PO Box 296, Epping, NSW 1710, Australia
\and
  Centre for Astrophysics Research, Science and Technology 
  Research Institute, University of Hertfordshire, Hatfield, AL10 9AB, UK    
\and
  School of Physics \& Astronomy, Queens Building, Cardiff University,
  The Parade, Cardiff CF24 3AA, UK
\and
  Institute of Astronomy, School of Physics,
  The University of Sydney, NSW 2006, Australia
\and
  Laboratorio de Astrof\'{i}sica Espacial y F\'{i}sica Fundamental,
  European Space Astronomy Center, P.O.\ Box 78, 
  E-28691 Villanueva de la Ca\~{n}ada, Madrid, Spain
}
\date{Received 15 October 2008}

\abstract{Optical and near-infrared (NIR) spectra are presented for
  \secondqsolong\ (hereafter \secondqso), a new redshift $z = 6.127
  \pm 0.004$ quasar discovered in the Third Data Release (DR3) of the
  UKIRT Infrared Deep Sky Survey (UKIDSS).  
  The source has $Y_\vega = 19.10 \pm 0.03$, corresponding to $\mabs =
  -27.12$, which is comparable to the absolute magnitudes of the
  $z\simeq6$ quasars discovered in the Sloan Digital Sky Survey (SDSS).  
  \secondqso\ was, in
  fact, registered by SDSS as a faint source with $z_\ab =
  20.13\pm0.12$, just below the signal--to--noise ratio limit of the
  SDSS high-redshift quasar survey.  The faint \zprime-band magnitude is a
  consequence of the weak \lya/\nv\ emission line, which has a
  rest-frame equivalent width of $\simm 20 \unit{\angstrom}$ and
  provides only a small boost to the \zprime-band flux.  Nevertheless,
  there is no evidence of a significant new population of high-redshift 
  quasars with weak emission lines from this UKIDSS-based search.
  The \lya\ optical depth to \secondqso\ is consistent with that measured
  towards similarly distant SDSS quasars, implying that results from
  optical- and NIR-selected quasars may be combined in studies of 
  cosmological reionization.  \\ 
  Also presented is a new NIR-spectrum of the previously discovered UKIDSS
  quasar \firstqsolong, which reveals the object to be a broad
  absorption line quasar. The new spectrum shows that the emission
  line initially identified as \lya\ is actually \nv, leading to a
  revised redshift of $z=\zfirstqso$, rather than $z=\zfirstqsoold$ as
  previously estimated.}


\maketitle


\section{Introduction}
\label{section:intro}

Since their discovery by \citet{Schmidt:1963} and
\citet{Hazard_etal:1963}, quasars have continued to be the most
revealing probes of the high-redshift Universe (\eg,
\citealt{Schneider:1999}).  Even though galaxies have been detected
out to greater distances (\eg, IOK-1, with a redshift of $z = 6.96$,
\citealt{Iye_etal:2006}) and gamma ray bursts may briefly be more
luminous (\eg, \citealt{Haislip_etal:2006}), the highest-redshift
quasars (\eg, SDSS~1148$+$5251 at $z = 6.42$, \citealt{Fan_etal:2003};
CFHQS~J2329$-$0301 at $z = 6.43$, \citealt{Willott_etal:2007}) are
more useful because they remain bright enough to be investigated in detail.
It has been possible to obtain high signal--to--noise ratio 
(\ston) spectra
for all the known $z \simeq 6$ quasars, robustly confirming the nature
of these sources, revealing their intrinsic properties (\eg,
\citealt{Walter_etal:2004, Venemans_etal:2007}) and, through
absorption, probing the intervening matter out to the quasars' redshifts.
The resultant measurements of the $z \simeq 6$ quasar population 
are also of interest, as they provide critical 
limits on the early structure formation scenarios 
(\eg, \citealt{Kurk_etal:2007}).
Probably the most dramatic discovery from these studies
is the marked increase
in the optical depth to neutral hydrogen at redshifts of $z \ga 5.7$
\citep{Becker_etal:2001,Fan_etal:2002,Fan_etal:2006b}.  The increase
in optical depth appears to represent the end of cosmological
reionization (\eg, \citealt{Barkana_Loeb:2001}), a conclusion
supported by the {\em{Wilkinson Microwave Anisotropy Probe}}
({\em{WMAP}}; \citealt{Bennett_etal:2003}) measurements of the cosmic
microwave background (\eg, \citealt{Dunkley_etal:2009}).  While a
consistent picture of reionization has emerged, the direct
measurements of this process are limited to the small number of $z
\simeq 6$ quasars known, and it is clear that the discovery of quasars
with $z \ga 7$ is vital to further progress in this field.

The majority of the known $z \simeq 6$ quasars have been identified by
looking for point--sources with very red optical colours in wide-field
surveys such as the Sloan Digital Sky Survey (SDSS;
\citealt{York_etal:2000}) and the Canada France High-$z$ Quasar Survey
(CFHQS; \citealt{Willott_etal:2007}), and more discoveries will be
made in such projects. Optical searches are, however, unlikely to
probe beyond the current redshift limits.  Almost all $z \simeq 6$
quasar emission at rest-frame wavelengths shorter than the \lya\ transition at
$\lambda = 0.1216\unit{\micron}$ is absorbed by intervening hydrogen,
and such quasars are effectively dark at observed wavelengths
below $\lambda \simeq [0.85 +
0.12 (z - 6)] \unit{\micron}$.  Conversely, most optical
charge--coupled device (CCD) detectors have a poor response beyond
wavelengths of $\lambda \simeq 1 \unit{\micron}$ (\ie, redward of
the \zprime\ or $Z$ bands), so quasars with a redshift of $z \ga 6.4$
are destined to remain invisible to CCD-based surveys.  Given the
rarity of $z \ga 6$ quasars
(\eg, a surface density of $\simm 0.02 \unit{deg}^{-2}$ to 
$\zprime_\ab = 21.0$;
\citealt{Jiang_etal:2007}), progress can only be made using wide-field
surveys at longer wavelengths.

In the long term, radio surveys with, \eg, the Low Frequency Array
(LOFAR\footnote{See the LOFAR web-site at
{\tt{http://www.lofar.org/}}.})  and the Square Kilometre Array
(SKA\footnote{See the SKA web-site at
{\tt{http://www.skatelescope.org/}}.})  will provide a powerful
complementary approach (\eg, \citealt{Wyithe_etal:2009}), but the
first steps beyond the current limits will come in the near-infrared
(NIR), with surveys following the same basic principles as the SDSS
and CFHQS searches.
The largest completed NIR survey, the Two Micron All Sky Survey
(2MASS; \citealt{Skrutskie_etal:2006}), with a magnitude limit of
$J_\vega \simeq 15.8$, does not have sufficient depth to find any
plausible high-redshift quasars.  The Visible and Infrared Survey
Telescope for Astronomy (VISTA; \citealt{Emerson_etal:2004}) should
cover $\simm 2 \times 10^4 \unit{\sqdeg}$ to $J_\vega \simeq 20$
during the next decade, but progress in the search for high-redshift
quasars will come first from the partially complete UKIRT Infrared
Deep Sky Survey (UKIDSS; \citealt{Lawrence_etal:2007}).

One new high-redshift quasar, \firstqsolong, hereafter \firstqso, with
an estimated redshift of $z = \zfirstqsoold$
\citep{Venemans_etal:2007}, has already been discovered in UKIDSS; 
the second such discovery, \secondqsolong, hereafter \secondqso, is
presented here.  \sect{ukidss} gives an introduction to the UKIDSS LAS
project and \sect{selection} describes the selection techniques that
led to \secondqso\ being identified as a candidate high-redshift
quasar.  Optical and NIR spectra of \secondqso\ are presented and
compared to those of $z \simeq 6$ SDSS quasars in \sect{secondqso}.
In addition, a new NIR spectrum of \firstqso\ is presented in
\sect{firstqso}, together with a revised redshift estimate for this
source.  The conclusions and future prospects for
high-redshift quasar searches with UKIDSS are discussed in
\sect{conc}.

All photometry is given in the native system of the telescope in
question, and explicitly subscripted.  
Thus SDSS \iprime\ and \zprime\ photometry is on the AB
system, whereas UKIDSS $Y$ and $J$ photometry is Vega-based.  The AB
corrections for the UKIDSS bands are $Y_{\rm{AB}} = Y_{\rm{Vega}} +
0.634$ and $J_{\rm{AB}} = J_{\rm{Vega}} + 0.938$
\citep{Hewett_etal:2006}.  Calculations of absolute (AB) magnitudes
are performed assuming a fiducial flat cosmological model with
normalised matter density $\Omega_{\rm{m}} = 0.27$, 
normalised vacuum density $\Omega_\Lambda = 0.73$, 
and Hubble constant $H_0 = 71 \unit{km} \unit{s}^{-1} \unit{Mpc}^{-1}$.


\section{The UKIRT Infrared Deep Sky Survey}
\label{section:ukidss}

UKIDSS \citep{Lawrence_etal:2007} is a suite of five surveys
undertaken with the Wide Field Camera (WFCAM;
\citealt{Casali_etal:2007}) on the 3.8\unit{m} United Kingdom Infrared
Telescope (UKIRT) at Mauna Kea, Hawaii.  The WFCAM detectors are four
sparse-packed $2048\times2048$ Rockwell Hawaii-II arrays, each of
which has a field of view of $0.05 \unit{\sqdeg}$.  Observations are
generally undertaken in sets of four contiguous pointings which,
together, cover $0.77 \unit{\sqdeg}$.  Full details of the survey
operations can be found in \citet{Dye_etal:2006}.  The individual
images are analysed by the data reduction pipeline described by
\citet{Irwin_etal:2009} and the catalogues of detected objects are
merged across bands into a queryable relational database at the WFCAM
Science Archive\footnote{The WSA is located at
{\tt{http://surveys.roe.ac.uk/wsa/}}.}  (WSA;
\citealt{Hambly_etal:2008}).
 
The UKIDSS Large Area Survey (LAS) will cover 4000\unit{\sqdeg} within
the SDSS footprint with a series of 40\unit{s}
exposures\footnote{80\unit{s} exposures are sometimes taken in
mediocre conditions, an example of which is the finding chart shown in
\fig{findingchart}.} in the $Y$, $J$, $H$ and $K$ bands.  The 
resulting magnitude limits are typically $Y_\vega \simeq 20.2$, $J_\vega
\simeq 19.6$, $H_\vega \simeq 18.8$ and $K_\vega \simeq 18.2$ for
point--sources detected with $\ston \simeq 5$,
and the average seeing is $0\farcs8$ \citep{Warren_etal:2007}.  
The area, depth and wavelength coverage of
the LAS were chosen with the detection of $z \simeq 6$ quasars in mind,
and the $Y$ filter, positioned between the $\zprime$ and $J$ bands in
a region of minimal atmospheric absorption ($0.97 \unit{\micron} \la
\lambda \la 1.07 \unit{\micron}$), is particularly useful in this
regard \citep{Hewett_etal:2006}.  Quasars in the redshift interval
$6.4 \la z \la 7.2$ would be identifiable as being unusually red in
\imy\ and \zmy, while being significantly bluer in \ymj\ than the
more numerous L and T dwarfs.

All UKIDSS images and catalogues are made available to European
Southern Observatory (ESO) countries and then, 18 months later, the
world, in a series of incremental data releases.  The results
presented here are based on the Third Data Release (DR3;
\citealt{Warren_etal:2009}) for which there is $870\unit{\sqdeg}$ of
LAS coverage in the $Y$ and $J$ bands.


\section{Candidate selection}
\label{section:selection}

High-redshift quasar candidates are selected in a two-step process,
using survey data initially (\sect{initial}), followed by additional
photometric observations of the most promising objects
(\sect{followup}).

\subsection{Initial shortlist}
\label{section:initial}

Redshift $\simm 6$ quasars are expected to appear as stationary
point--sources with extremely red optical--NIR colours.  The first
stage of the candidate selection process is to extract a fairly
complete (if highly contaminated) sample of all such sources from the
UKIDSS and SDSS databases.

The initial selection from the UKIDSS LAS was of all point--sources
with $Y_\vega \leq 19.88$ that were also detected in the $J$ band and
have measured ${\mbox{$Y_\vega$$\!\,-$$J_\vega$}} \leq 0.88$ (to
exclude L and T dwarfs).  The resulting sample was then cross-matched
to the SDSS Fifth Data Release (DR5;
\citealt{AdelmanMcCarthy_etal:2007}) and all sources either undetected
by SDSS or with ${\mbox{$i_\ab$$\!\,-$$Y_\vega$}} \geq 2.5$ (but
undetected in \uprime, \gprime\ and \rprime) were selected.  Any
sources with inter-band (or inter-survey) positional mis-matches of
greater than $0\farcs7$ were rejected to avoid nearby Galactic
stars with appreciable proper motions, as well as most
asteroids\footnote{Some asteroids, observed at a turning point between
prograde and retrograde motion, can appear stationary in UKIDSS survey
images taken within an hour of each other.}.

A number of heuristic algorithms were applied to reject sources for
which the UKIDSS or SDSS database photometry is likely to be
unreliable.  Contaminating sources eliminated at this stage include
those arising from WFCAM cross-talk \citep{Dye_etal:2006},
objects in the haloes of bright stars, and close pairs separated by a
few arcsec for which deblending is required in SDSS.

For the sources that were undetected in SDSS \iprime\ and \zprime,
aperture fluxes, corrected for aperture loss, were measured from the
SDSS images.  Acquisition of the aperture fluxes represents the final
step in a fully automatic procedure that, in the analysis of UKIDSS
DR3, yielded $\simm 2 \times 10^4$ pre-candidates with \iprime-, \zprime-,
$Y$- and $J$-band photometry.

The next stage of the selection process is to use 
Bayesian model comparison to determine the
probability,
given the photometric data, 
that each source is a high-redshift quasar, $\probqso$.  
This calculation is described in detail in \citet{Mortlock_etal:2009c}.  
The key to this approach is having an accurate model of the stellar 
population, specifically the intrinsic magnitude and colour distributions
of the cool M dwarfs
which scatter into the quasar selection box described above.
Convolving this distribution in \iprime,
\zprime, $Y$ and $J$ with the observational noise gives the likelihood
that any candidate is an M dwarf.  
This can be compared directly with the likelihood that the candidate is 
a quasar, which is calculated 
using an extrapolation of the high-redshift quasar luminosity function
of \citet{Fan_etal:2004} and simulated colours from \cite{Hewett_etal:2006}.
The result is a fully self-consistent value
of $\probqso$ that, in principle, combines the available information on the
candidate with the constraints on the quasar and star populations
in an optimal way.
Of course there are practical limitations, the most important of
which result from incomplete sampling of the tails of the noise 
distributions and the limited knowledge of the
stellar population fainter than the UKIDSS and
SDSS survey limits.  The latter is important as the probability calculation
necessarily includes the possibility of faint sources scattered into
the sample, but is difficult to assess empirically using the survey data.
Despite these ambiguities 
(which are explored further in \citealt{Mortlock_etal:2009c}),
the probabilities which result from the adopted noise and population models
clearly provide a good objective ranking scheme for quasar candidates.

Most of the candidates are clustered close to the stellar locus and
their low quasar probabilities (\ie, $\probqso \la 0.01$) merely
confirm the obvious.  The motivation for the method is to
assess the ambiguous sources lying between the quasar
and star loci in colour space, or close to the $Y$-band limit
with quasar-like colours but large errors.  The sometimes
counter-intuitive results are explored in some detail in
\citet{Mortlock_etal:2009c}, but the critical point is that the
candidates can be ranked by $\probqso$ and thus prioritised
objectively for further investigation.

\begin{figure}
\centering
\includegraphics[width=\figwidth, angle=0]{\figdir 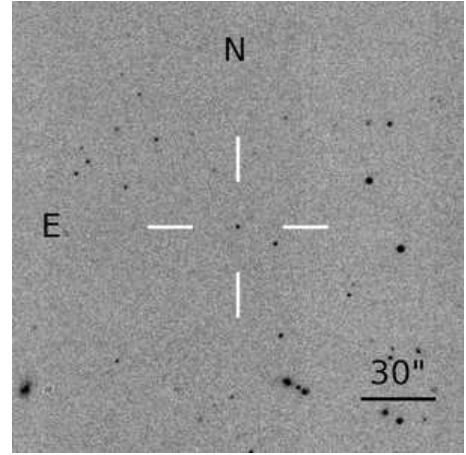}
\caption{Finding chart for \secondqsolong\ showing the UKIDSS LAS
  $80\unit{s}$ $J$-band image of a field $3\unit{arcmin}$ on a side
  centred on the source position at ${\rm{RA}} = 199.79706\unit{deg}$
  and ${\rm{dec}} = 9.84762\unit{deg}$.}
\label{figure:findingchart}
\end{figure}

It is only at this stage that any visual inspection was undertaken,
with the UKIDSS and SDSS images of the few hundred best candidates
checked for artefacts such as hot pixels, bad columns, 
obviously wrong photometry, 
and 
undetected, blended or moving sources.
(Although the UKIDSS and SDSS data reduction pipelines flag the 
vast majority of such instances, the selection of unusual sources 
on the basis of observed inter-survey colours inevitably 
produces samples with an over-representation of the very rare cases
for which at least one of the surveys has produced anomalous measurements.)
Following visual inspection, 
only a small sample of $\la 100$ plausible quasar candidates
with $\probqso \ga 0.01$ remained\footnote{The choice of
threshold value for $\probqso$ (which defines the list of candidates
and determines the completeness and contamination of the quasar sample)
depends on the resources available for follow-up photometry and
spectroscopy.}, one of which was \secondqso, shown in \fig{findingchart}.
Follow-up observations were needed to 
either reject these candidates as stars or, 
hopefully, confirm some as high-redshift quasars.  

\subsection{Follow-up photometry}
\label{section:followup}

Rather than immediately taking spectra of all the candidates (\cf\
\citealt{Glikman_etal:2008}), it is more efficient first to refine the
photometry, obtaining short exposures in the \iprime, \zprime, $Y$ and
$J$ bands on a variety of telescopes.  The quasar probability defined
in \sect{initial} is recalculated whenever new data are obtained, and
a source is dropped from the candidate list if $\probqso$ falls
below the selection threshold at any point.  
As almost all the candidates initially
have minimal \ston\
in the \iprime\ band SDSS images, follow-up
observations in this band are the most efficient way to quickly reject
such sources, revealing most to be cool stars scattered to have
quasar-like colour by photometric noise.  Observations in \iprime\ are
thus prioritised, although initial follow-up observations are
sometimes made in \zprime, $Y$ or $J$, depending on weather,
telescope scheduling, and other external factors. Candidates are not
considered for spectroscopy until follow-up photometry 
has been obtained in at least the $\iprime$, $Y$ and $J$ bands.

Follow-up observations of \secondqso\ were obtained in \iprime,
\zprime, $Y$ and $J$, as summarised in \tabl{UJphot}.  

\secondqso\ was
first re-observed in the \iprime\ band at the Liverpool Telescope (LT)
on the nights beginning 2008 January 8 and 12 for a total of
$3240\unit{s}$.  The candidate's quasar probability remained high, and
so $300 \unit{s}$ exposures in each of the $Y$ and $J$ bands were
obtained using the UKIRT Fast-Track Imager (UFTI) on the night
beginning 2008 January 16.  The candidate still appeared promising so
the UFTI observations were repeated the following night.

The improved \iprime-, $Y$- and $J$-band measurements were
sufficiently precise that \secondqso\ had $\probqso \simeq 1$
(\ie, it was essentially implausible for a cool star to have scattered
to the candidate's observed colours given the precision of the new
photometric data), whereas all the other initial candidates were
rejected on the basis of their follow-up photometry. 
A spectrum of \secondqso\ was thus 
obtained on the night beginning 2008 January 22,
with the result that it was immediately confirmed 
as a high-redshift quasar (\sect{secondqso}).  
That the entire selection, follow-up and
confirmation process for \secondqso\ was completed less than seven
weeks after the UKIDSS DR3 release (on 2007 December 6) is a powerful
illustration of the advantages of queue-scheduling for the three
telescopes involved, especially considering the RA of the target and
the time of year.

\secondqso\ had already been detected as a faint source in SDSS,
although with $\zprime_\ab = 20.13 \pm 0.12$ its \ston\
was too low for it to be selected into the high-redshift quasar sample 
defined by \cite{Fan_etal:2003}.  Post-confirmation, improved \zprime-band
photometry of \secondqso\ was obtained using the ESO Multi-Mode
Instrument (EMMI) on the New Technology Telescope (NTT) on the night
of 2008 January 29.  The observations were made with the long-pass
$\#611$ filter, the bandpass of which, in combination with the red
cut-off of the CCD response, is quite similar to the SDSS \zprime\
bandpass \citep{Venemans_etal:2007}.  To calibrate the NTT-image, SDSS
\iprime- and \zprime-band photometry of bright, unsaturated stars in
the frame was converted to the natural system of the image using
$z_{\rm{NTT,AB}} = \zprime_\ab - 0.05 (\iprime_\ab - \zprime_\ab)$
\citep{Venemans_etal:2007}, and the result quoted in \tabl{UJphot} is
in this natural system.

\begin{table*}
\center
\begin{minipage}{121mm}
\centering
\caption{Original survey and follow-up photometric observations of \secondqso.}
\begin{tabular}{lccllr}
\hline \centering filter & original & follow-up 
  & observing date & telescope & exposure time \\ 
\hline 
$\iprime_\ab$ & 22.83$\pm$0.32 & 22.55$\pm$0.09 
  & 2008 January 8 and 12 & LT & $2 \times 1620\unit{s}$ \\ 
$\zprime_\ab$ & 20.13$\pm$0.12 & 19.99$\pm$0.03 
  & 2008 January 29 & NTT & 450\unit{s} \\ 
$Y_\vega$ & 19.22$\pm$0.06 & 19.10$\pm$0.03 
  & 2008 January 16 and 17 & UKIRT & $2 \times 300\unit{s}$ \\ 
$J_\vega$ & 18.69$\pm$0.05 & 18.76$\pm$0.03 
  & 2008 January 16 and 17 & UKIRT & $2 \times 300\unit{s}$ \\ 
\hline
\end{tabular}
All magnitudes are quoted in the natural system of the
  initial survey: AB for the optical SDSS bands and Vega for the NIR
  UKIDSS bands.  Note that the transmission profile of the
  $\zprime_{\rm{NTT}}$ filter differs somewhat from that of the SDSS
  \zprime\ filter (see \sect{followup}).
\label{table:UJphot}
\end{minipage}
\end{table*}


\section{\secondqso}
\label{section:secondqso}

After the series of follow-up photometric observations described in
\sect{followup} showed \secondqso\ to be a promising high-redshift quasar
candidate, an optical spectrum was obtained to confirm the
identification (\sect{spectra}). A more accurate redshift was
estimated from a NIR spectrum covering the \mgii\ emission line
(\sect{redshift}), after which both spectra were used to compare
\secondqso 's emission and absorption properties with those of
similarly distant quasars discovered in SDSS (\sect{sdsscomparison}).

\subsection{Spectroscopic observations}
\label{section:spectra}

\begin{figure}
\centering
\includegraphics[width=\figwidth, angle=90]{\figdir 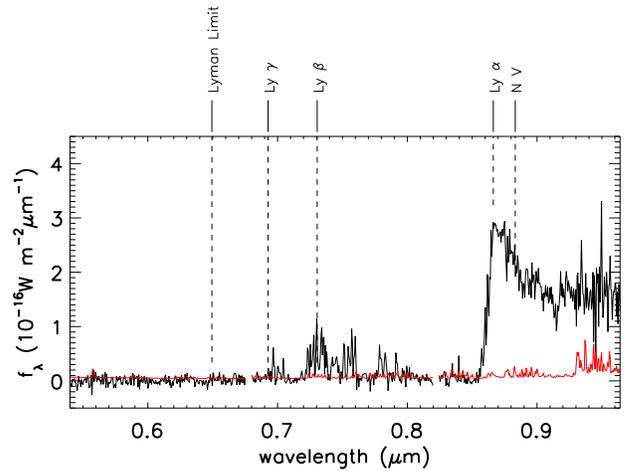}
\caption{The Gemini GMOS spectrum of \secondqso\ (black curve) and the
  noise spectrum (red curve), both binned by a factor of four.  
  The gaps at wavelengths of $\simm 0.68 \unit{\micron}$
  and $\simm 0.82\unit{\micron}$ correspond to the breaks between
  the different CCDs.
  The wavelengths of common emission lines redshifted by $z = 6.13$ are
  indicated.}
\label{figure:gmospec}
\end{figure}

An optical spectrum of \secondqso\ was obtained using the Gemini
Multi-Object Spectrograph (GMOS) on the
Gemini South Telescope on the night beginning 2008 January 22.  Two
spatially-offset spectra, each of duration 900\unit{s}, were obtained
using a $1\farcs0$ slit and the R400 grating, covering
the wavelength range of 0.5--1.0\unit{\micron} over the three CCDs.
The standard bias subtraction and flat-fielding steps were followed.
Then, because of the strong sky lines in the red part of the spectrum,
the `double subtraction' method was used for the first-order sky
subtraction (\ie, frame B was subtracted from frame A, and then the
negative spectrum subtracted from the positive spectrum, after the two
spectra were aligned). This procedure removes most systematic errors,
but at the price of increasing the noise in the final frame by a
factor of $\simm1.4$ compared to the theoretical limit.
Second-order sky subtraction was achieved by fitting a smooth function
to each column. At this point cosmic rays were removed using the
Laplacian Cosmic Ray Removal Algorithm (LCRRA;
\citealt{VanDokkum:2001}).  Wavelength-calibration was carried out
using observations of a Cu\,Ar lamp.  Relative spectrophotometry, and
correction for telluric absorption, was achieved using observations of
a standard star, and the spectrum was then scaled to match the NTT
\zprime-band photometry given in \tabl{UJphot}.  The final GMOS
spectrum is shown in \fig{gmospec}.

The spectrum is recognisable as that of a $z \simeq 6$ quasar from the
presence of a broad emission line, identified as \lya, at the same
wavelength as a strong continuum break, attributed to \lya\ forest absorption.
The \lya\ emission peaks at a wavelength of $\lambda \simeq 0.88
\unit{\micron}$, which leads to a preliminary redshift estimate of 
$z = 6.12$.  However the \lya\ line is strongly absorbed to the blue and
such high-ionization lines can exhibit velocity shifts relative to the
quasar systemic redshift (\eg, \citealt{Tytler_Fan:1992}), 
limiting the utility of the optical redshift measurement.

\begin{figure}
\centering
\includegraphics[width=\figwidth, angle=90]{\figdir 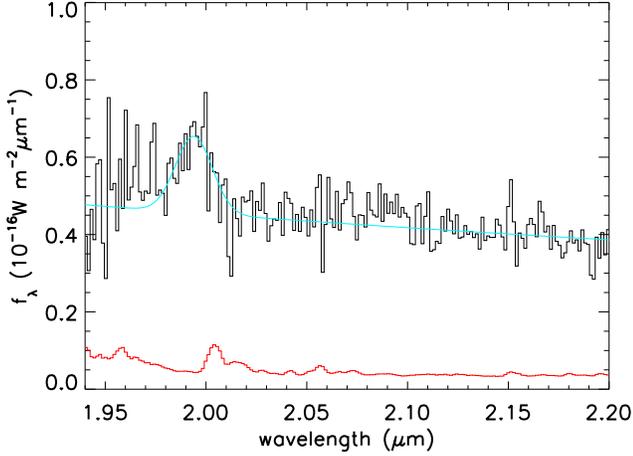}
\caption{The Gemini NIRI spectrum of \secondqso\ (black curve) and the
error spectrum (red curve), both binned by a factor of two.  The
continuum+line fit to the \mgii\ emission line is also shown (blue).
The fit implies a source redshift of $z = 6.127 \pm 0.004$.
The effect of telluric absorption was removed using a standard star
observed at similar airmass, resulting in the increased errors at
$\simm2.005\unit{\micron}$.}
\label{figure:mgiifit}
\end{figure}

The \mgii\ line, if observable in the $K$ band, should give a more reliable
estimate of the redshift, and so a NIR spectrum of \secondqso\ was
obtained.  The source was observed using the Near-IR Instrument (NIRI)
on the Gemini North Telescope on the two nights beginning 2008
February 26 and 27. Observations were made with a $0\farcs75$ slit
and the $K$ grism G5204, covering the wavelength range
1.9--2.5\unit{\micron}
with a resolving power of $R = 500$.
With NIRI it is standard procedure to discard
the first exposure of a sequence, leaving a total of 11 usable
300\unit{s} exposures over the two nights.  The observation and data
reduction methodology of \citet{Weatherley_etal:2005} was adopted,
with six different slit positions used to reduce the noise from
sky-subtraction relative to the more standard ABBA sequence. The data
suffered from increasing slit losses as the observations proceeded,
which is believed to be due to differential flexure between the guide
probe and the instrument. The additional slit losses, relative to the
first frames on each night, reduced the final \ston\ of the detected
emission line by a factor of $\simm1.5$. Because of the varying
throughput, each sky-subtracted frame was first scaled to a common
count level, and then weighted by the inverse variance in the sky (as
evaluated in a region free of strong emission lines).  Wavelength
calibration was performed using the list of sky emission lines from
\citet{Rousselot_etal:2000}. A standard star observed at similar
airmass was used to correct for telluric absorption, and for relative
flux calibration, after which the spectrum was scaled to match the
UKIDSS $K$-band photometry given in \tabl{UJphot}.  The final NIR
spectrum is plotted in \fig{mgiifit}.

\subsection{Redshift estimation}
\label{section:redshift}

\begin{figure}
\centering
\includegraphics[width=\figwidth, angle=90]{\figdir 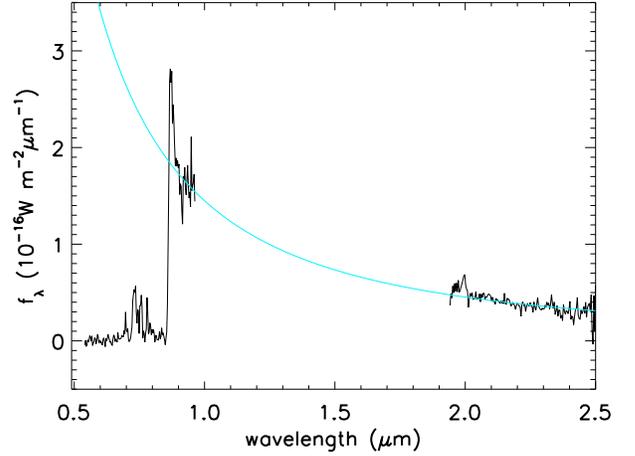}
\caption{The optical spectrum, binned by a factor of four (left black
  curve), and the NIR spectrum, binned by a factor of five (right
  black curve), of \secondqso.  The power-law continuum fit is also
  shown (blue).}
\label{figure:totalspec}
\end{figure}

The NIR spectrum of \secondqso\ shown in \fig{mgiifit} reveals the
broad \mgii\ emission line near $\lambda \simeq 2.0 \unit{\micron}$,
corroborating the initial redshift estimate of 
$z \simeq 6.1$.  To obtain a more accurate
measurement (and to quantify the absorption in the
\lya\ forest in \sect{sdsscomparison}) a power-law fit to the
continuum was made to the combined optical and NIR spectrum by
minimizing $\chi^2$ and iteratively clipping outliers to eliminate the
emission lines.  The best-fit power-law is $f_\lambda (\lambda)
\propto \lambda^{-(2 - \alpha)}$, where $\alpha = 0.32 \pm 0.01$ (
defined so that $f_\nu \propto \nu^{-\alpha}$). This is plotted in
\fig{totalspec}.  This continuum was subtracted from the data and then
a Gaussian was fit to the residual \mgii\ emission line; the resultant
continuum+line fit is plotted in \fig{mgiifit}.  The central
wavelength of the \mgii\ emission is $\lambda = (1.994 \pm 0.001)
\unit{\micron}$, implying that \secondqso\ has a redshift of
$z=6.127\pm0.004$.  The \mgii\ emission line has a rest-frame
equivalent width of $\simm 14\unit{\angstrom}$, substantially less
than the typical value of $\simm 28\unit{\angstrom}$ for $z \simeq 6$
quasars found by \cite{Kurk_etal:2007}.

\subsection{Comparison with SDSS quasars}
\label{section:sdsscomparison}

\secondqso\ is, after \firstqso\ \citep{Venemans_etal:2007}, only the
second redshift $\simm6$ quasar discovered in the NIR.  Given that
surveys at these wavelengths are the first capable of probing beyond
the current limit of $z \simeq 6.4$, it is important to determine
whether quasars selected in the NIR exhibit significant
differences from those selected in the optical.  
This preliminary exploration focuses on 
the emission line strengths, 
the neutral hydrogen optical depth, 
and the ionization region around the quasar.

\subsubsection{Emission line properties}

\begin{figure}
\centering
\includegraphics[width=\figwidth, angle=90]{\figdir 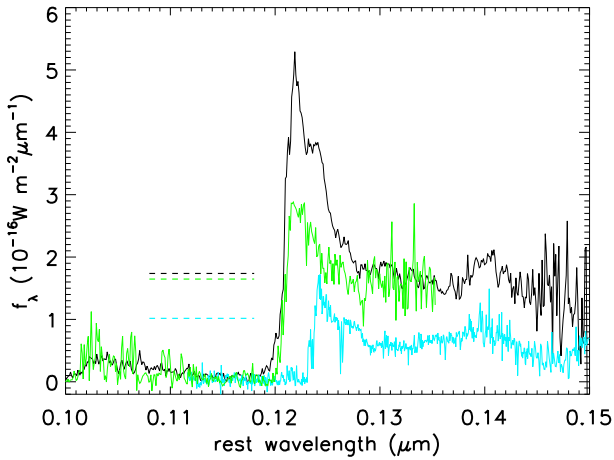}
\caption{Rest-frame spectra of \firstqso\ (blue) and \secondqso\
  (green) compared to the composite spectrum of $z \simeq 6$ SDSS
  quasars presented by \cite{Fan_etal:2004}.  
  The continuum level adopted for the calculation of the equivalent with 
  of the \lya\ line for each spectrum is also shown (dashed lines).}
\label{figure:qcompare}
\end{figure}

Redshift $\simm 6$ quasars are identifiable in photometric surveys
primarily due to their extreme colours which result from the strong
neutral hydrogen absorption blueward of the quasars' \lya\ emission
lines.  However \fig{qcompare}, which compares the rest-frame spectrum
of \secondqso\ (and \firstqso; see \sect{firstqso}) to a composite
SDSS quasar spectrum \citep{Fan_etal:2006c}, shows that \secondqso\
has a noticeably weaker \lya\ emission line than is typical for
optically-identified sources.

Measuring the
rest-frame equivalent width of \secondqso's \lya\ line is somewhat
problematic as the continuum level is ambiguous and the power-law fit
shown in \fig{totalspec} is probably too steep to give a good local
continuum estimate.  For the purposes of comparison the continuum
levels shown in \fig{qcompare} were adopted, yielding rest-frame
equivalent widths of $19.3 \unit{\angstrom}$ for \secondqso\ and 
$60.2 \unit{\angstrom}$ for the \cite{Fan_etal:2004} composite.
This quantifies 
the visual impression that \secondqso\ has a weaker \lya\
emission line than an average $z \simeq 6$ quasar, 
although there are several examples of high-redshift quasars
with even weaker lines
(\eg, \citealt{Fan_etal:1999b,DiamondStanic_etal:2009}).

If \secondqso\ differed only in that its \lya\ emission was marginally
stronger (or, equivalently, less absorbed to the blue), it would
almost certainly have been discovered in the SDSS high-redshift quasar 
sample.
With $z_\ab = 20.13 \pm 0.12$ in the SDSS database, it was brighter
than the \cite{Fan_etal:2003} limit of $\zprime_\ab = 20.2$, but it
failed the \ston\ cut of $\sigma_{z_\ab} < 0.1$.  Given that the NTT
re-observation described in \sect{followup} gave $\zprime_\ab = 19.99
\pm 0.03$, the SDSS team were unlucky that the measured \zprime-band flux was
less than the true value and that the source was observed in slightly
worse than average conditions, both of which combined to leave the
source outside their selection cuts. This
illustrates the more general limitation inherent in any selection
method which applies a hard data cut in a region of parameter space
where there are appreciable numbers of objects.  A way to avoid the
problem, at least in principle, is to apply the probabilistic approach
adopted in \sect{initial} and described in full by
\cite{Mortlock_etal:2009c}, although it remains difficult to deal with
the large numbers of low \ston\ candidates near the survey limit.

\subsubsection{Neutral hydrogen optical depth}

Observations of $z \simeq 6$ quasars have revealed a marked increase
in the density of neutral hydrogen above a redshift of $\simm 5.7$
(\eg, \citealt{Fan_etal:2002}), possibly indicating the end of
cosmological reionization (\eg, \citealt{Fan_etal:2006b}).  As
hydrogen absorption also has an effect on the detectability of $z
\simeq 6$ quasars, it is important to assess whether the measured
optical depths of optical- and NIR-selected quasars differ
significantly.

Here, the spectrum of \secondqso\ is analysed following the method of
\cite{Fan_etal:2006b}, with the effective optical depth,
$\taueff(\zabs)$, at redshift $\zabs$ estimated to be
\begin{equation}
\taueff(\zabs) = - \ln \left\{ \frac {\langle\fluxobs[(1 + z)
 \lambda_\lya] \rangle_{z \simeq \zabs}} {\langle \fluxintr[(1 + z)
 \lambda_\lya] \rangle_{z \simeq \zabs}} \right\},
\end{equation}
where $\fluxobs(\lambda)$ is the observed flux density,
$\fluxintr(\lambda)$ is the continuum flux density (assumed to be
given by the power-law fit described in \sect{redshift}), and
$\lambda_{\lya} = 0.1216\unit{\micron}$ is the rest-frame wavelength
of the \lya\ transition.  The angle brackets denote an averaging over
a finite redshift range around $\zabs$;
bins of width $\Delta \zabs = 0.15$ were used.  
Given that the \lyb\
and \lya\ emission lines appear at $\lambda = 0.74\unit{\micron}$ and
$\lambda = 0.87\unit{\micron}$, respectively, $\taueff(\zabs)$ can be
estimated in the range $5.1 \la \zabs \la 6.0$.  The errors in these
optical depth estimates are dominated by the noise in the spectrum of
\secondqso, with the uncertainties in the continuum fit and sample
variance (across different lines--of--sight) being secondary.  
The resultant error bars are also significantly asymmetric,
especially for the $\zabs = 5.93$ bin, in 
which flux is detected with only $\ston \simeq 2$.
These \lya\ optical depth estimates, along with an analogous
estimate of the \lyb\ optical depth (with the \lya\ emission corrected
for as described by \citealt{Fan_etal:2006b}), are given in \tabl{optdepth}.

These estimated optical depths towards \secondqso\ are consistent with
measurements of SDSS quasars by \cite{Fan_etal:2006b}, 
which gives some confidence that
\cite{Gunn_Peterson:1965} effect measurements of optical- and
NIR-selected $z \simeq 6$ quasars may be combined.

\begin{table}
\center
\begin{minipage}{41mm}
\centering
\caption{Effective optical depths of the absorption
  systems seen in the spectrum of \secondqso.}
\begin{tabular}{lccc}
\hline transition & $\zabs$ & $\taueff$ \\
\hline 
\lya & 5.93 & $3.59_{-0.41}^{+0.71}$ \\
\lya & 5.78 & $4.27_{-0.31}^{+0.45}$ \\
\lya & 5.63 & $3.35_{-0.11}^{+0.13}$ \\
\lya & 5.48 & $2.58_{-0.12}^{+0.14}$ \\
\lya & 5.33 & $3.21_{-0.13}^{+0.15}$ \\
\lya & 5.18 & $1.94_{-0.03}^{+0.03}$ \\
& & \\ 
\lyb & 5.88 & $2.43_{-0.06}^{+0.06}$ \\
\hline
\end{tabular}
The \lyb\ optical depth
  has been corrected for \lya\ absorption using the relation found by
  \citet{Fan_etal:2006b}.
\label{table:optdepth}
\end{minipage}
\end{table}

\subsection{Quasar ionization region}
\label{section:stromgren}

The ultraviolet (UV) photons from quasars are sufficiently energetic
to ionize the neutral hydrogen in a volume around them, creating a
\cite{Stromgren:1939} sphere.  The extent of the ionization region
depends in part on the UV luminosity of the quasar and the
neutral fraction of the surrounding medium, potentially providing a
complementary probe of reionization.  Unfortunately, the measurement
is difficult in practice, due to uncertainties in the quasars' UV
spectral energy distributions (SEDs), unknown ionizing flux from
nearby galaxies, and difficulty in establishing the extent of the
\hii\ region from heavily absorbed and otherwise noisy spectra
(\eg, \citealt{Fan_etal:2006b}).

It has proved more useful to adopt a relative approach in which the
evolution of the size of quasars' \hii\ regions with redshift is
estimated without reference to an absolute model of the neutral
hydrogen fraction.  The relative measurement is attempted here for
\secondqso, once again following the methodology of
\cite{Fan_etal:2006b}, by defining the quasar proximity region as the
volume within which the transmitted flux fraction is $\la 0.1$.  After
dividing the Gemini spectrum by the continuum+line fit (described in
\sect{redshift}) to convert to transmitted flux, the spectrum was
smoothed to a resolution of $20\unit{\angstrom}$.  This smoothed
transmission spectrum remains below 0.1 down to a wavelength of
$\lambda \simeq 0.856\unit{\micron}$; converting to redshift and then
to a co-moving physical length implies that \secondqso\ has a
proximity zone of radius $R_{\rm{p}} \simeq 5.1\unit{Mpc}$.

The inferred radius is broadly consistent with measurements from
optically-detected SDSS quasars \citep{Fan_etal:2006b}, although to
make a more quantitative comparison it is necessary to standardise the
above measurement of $R_{\rm{p}}$ to a fiducial absolute magnitude
$\mabs = -27$ according to $R_{{\rm{p}},27} = 10^{0.4(\mabs + 27)}
R_{\rm{p}}$.  With $\mabs = -27.12$ this gives $R_{{\rm{p}},27} \simeq
4.9\unit{Mpc}$ for \secondqso. This is lower than the 
fiducial value of
$\simm7\unit{Mpc}$ given by \cite{Fan_etal:2006b} for $z \simeq 6.1$ quasars,
although this difference is not significant given the 
range of $R_{{\rm{p}},27}$ values seen in the optical sample.


\section{\firstqso}
\label{section:firstqso}

\firstqso\ was the first $z\simeq6$ quasar found in UKIDSS,
\cite{Venemans_etal:2007} describing the discovery and reporting a
redshift of $z=\zfirstqsoold$. The source was subsequently recovered 
by \citet{Jiang_etal:2008} in their
survey for high-redshift quasars using deep co-added data from
multiple scans of SDSS Stripe 82;
they quote a redshift of $z=5.85$.
An improved redshift for \firstqso\ was
sought by acquiring a NIR spectrum (\sect{nirspec}).  These data
revealed \firstqso\ to be a broad-absorption line (BAL) quasar,
resulting in a modified redshift estimate (\sect{redshift_ven}), and
also showing some discrepancies compared to the existing optical
spectrum in the region 0.9--1.0\unit{\micron}. The origin of the
discrepancy was traced to the relatively poor quality of the published
standard star data used to flux-calibrate the optical spectrum, 
which has hence been re-reduced (\sect{nirspec}).

\subsection{Spectroscopic observations}
\label{section:nirspec}

\subsubsection{Optical spectrum}

\begin{figure}
\centering
\includegraphics[width=\figwidth, angle=90]{\figdir 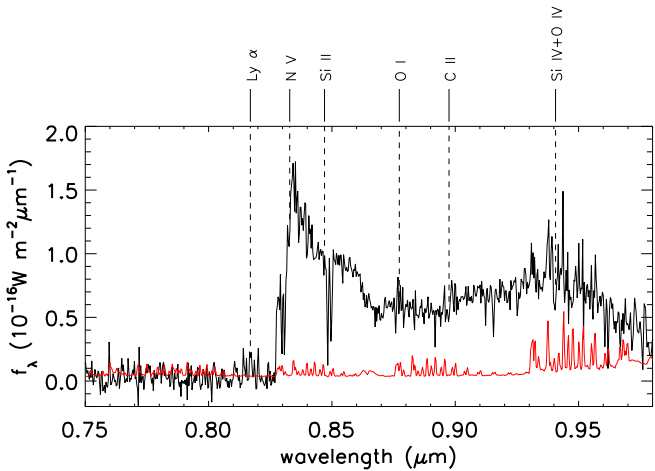}
\caption{Re-calibrated VLT spectrum of \firstqso\ (black curve) and
  the noise spectrum (red curve), binned by a factor of two.  The
  wavelengths of common emission lines redshifted by $z = \zfirstqso$
  are indicated.}
\label{figure:venspec}
\end{figure}

The original optical spectroscopic observations of \firstqso\ were made
in 2006 September using the
FOcal Reducer and low dispersion Spectrograph (FORS2) 
on the Very Large Telescope (VLT),
and are described in detail in \citet{Venemans_etal:2007}. 
The re-reduction followed a similar
sequence to that described there, up to the point of flux
calibration.  Instead of using the published
spectrophotometry of the white dwarf standard star GD50 from
\citet{Oke:1990}, a model spectrum with effective temperature
$T_{\rm{eff}} = 41150\unit{K}$ and surface gravity 
$\log[g / (1 \unit{cm} \unit{s}^{-2})] = 9.15$ from \cite{Dobbie_etal:2005} 
was used to establish the flux calibration curve.  Finally, the data were
scaled to match the measured $z$-band photometry. The revised
spectrum, plotted in \fig{venspec}, shows a sharp downturn longward of
0.95\unit{\micron} compared to the original spectrum.

\subsubsection{NIR spectrum}

A NIR spectrum of \firstqso\ was obtained using the Gemini Near
Infra-Red Spectrograph (GNIRS) on the Gemini South Telescope on the
night beginning 2007 March 22.  The observations were taken in
cross-dispersed mode with the short camera, the 32 lines mm$^{-1}$
grism, and a $0\farcs75$ slit, providing coverage from
$\simm0.8\unit{\micron}$ to $\simm2.5 \unit{\micron}$ at a resolving
power of $R=500$.  Due to the limited slit length it was impractical
to adopt the observation strategy used for the NIRI observations of
\secondqso\ (\sect{spectra}) and so a standard ABBA offset pattern was
used.

The observations comprise eight frames, with a total exposure time of
$2400\unit{s}$.  The data were reduced mainly using the Gemini--GNIRS
package in the Image Reduction and Analysis Facility (IRAF;
\citealt{Tody:1993}). After correction for pattern noise and
flat-fielding, each of the diffraction orders was separated, and the
double-subtraction method (described in \sect{spectra}) was applied to
each of the four pairs of images. The resulting four frames for each
order were then averaged.  S-distortion correction and wavelength
calibration were then applied using observations of an Ar lamp, and
one-dimensional spectra were extracted.  The data were corrected for
telluric absorption and flux-calibrated using the spectrum of a
spectroscopic standard star.  Finally, the different diffraction orders 
were spliced together to produce a single continuous spectrum.  Strong
telluric absorption bands occur at 1.35--1.43\unit{\micron} and
1.80--1.95\unit{\micron}, and so the data in these wavelength ranges
were discarded. The spectrum was placed on an absolute flux scale by
matching to the $K$-band photometry.

\begin{figure}
\centering
\includegraphics[width=\figwidth, angle=90]{\figdir 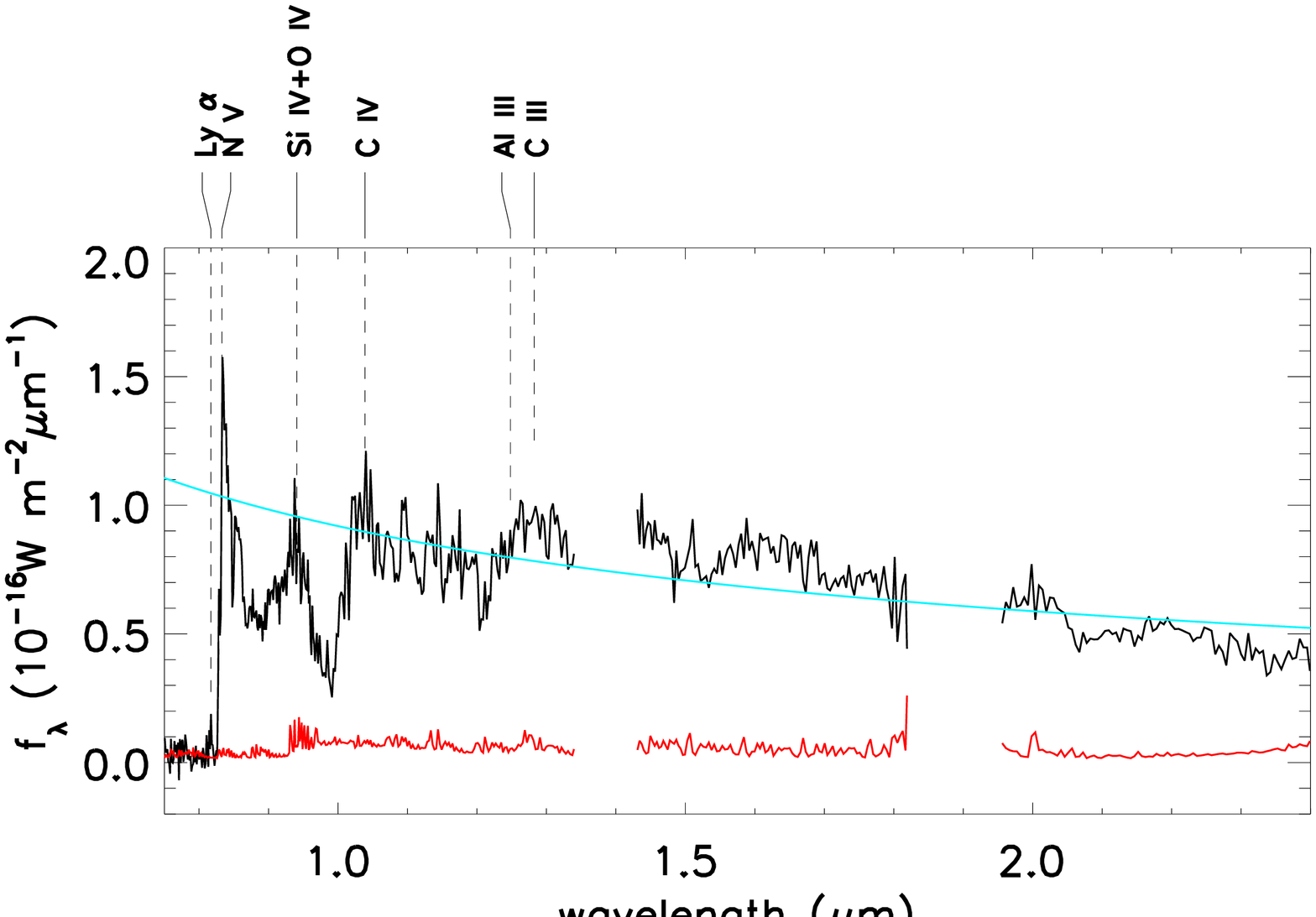}
\caption{The combined optical and NIR spectrum, spliced at $\lambda =
  0.96\unit{\micron}$, of \firstqso\ (black curve) and the noise
  spectrum (red curve), both binned by a factor of eight.  The
  power-law continuum fit is also shown (blue).  The wavelengths of
  common emission lines redshifted by $z = \zfirstqso$ are indicated.
  The \mgii\ line would be expected to appear at 1.88\unit{\micron},
  in the middle of the strong telluric absorption band at 
  $\simm 1.9\unit{\micron}$ for which the data have been omitted.}
\label{figure:full_venspec}
\end{figure}

Comparison of the optical and NIR spectra in the region of overlap
(0.8--1.0\unit{\micron}) showed good detailed agreement, except for a 
10\percent\ normalisation offset, and so the optical spectrum
was scaled to the NIR spectrum.  The declining \ston\ of the NIR spectrum
towards the blue, and of the optical spectrum towards the red, are
approximately equal at $0.96\unit{\micron}$, so the two spectra were
spliced at this point. The final combined spectrum is plotted in
\fig{full_venspec}.

\subsection{Redshift estimation}
\label{section:redshift_ven}

The most striking features of the combined \firstqso\ spectrum shown
in \fig{full_venspec} are the broad absorption lines near
0.9\unit{\micron} and 1.0\unit{\micron}.
These are attributed to \siiv\ and \civ, respectively. 
Despite the increased wavelength coverage, the redshift
is difficult to determine accurately, as is common with BAL quasars
(\eg\ \citealt{Trump_etal:2006}).  \cite{Venemans_etal:2007} and
\cite{Jiang_etal:2008} identified the peak of the line near
$0.835\unit{\micron}$ with \lya, and the shoulder at
$0.855\unit{\micron}$ with \nv, but matching \siiv,
\civ\ and \ciii\ lines for this redshift are not evident. Instead it seems
likely that the peak at $0.835\unit{\micron}$ is the \nv\ emission
line, and that the \lya\ emission line has been absorbed by a \nv\
BAL. As shown in \fig{full_venspec}, a redshift of $z=\zfirstqso$
provides a reasonable match to the \nv\ peak, as well as weak features
that match \siiv, \civ, and \ciii; but any redshift in the range
$5.70<z<5.74$ is consistent with the data. 
Unfortunately the \mgii\ line,
which ought to give a definitive redshift, 
lies in the region of atmospheric absorption near $1.9\unit{\micron}$.

In the wavelength region 0.85--1.08\unit{\micron} it is difficult to
determine the extent of the BAL troughs, and the regions of unabsorbed
continuum emission.  The best fit power-law, plotted in
\fig{full_venspec}, is a poor match. It seems likely that the sharp
step near 0.865\unit{\micron} marks the blue edge of the \siiv\ BAL,
and that the shoulder at 0.855\unit{\micron}, previously identified
with \nv, is in fact continuum emission. The resulting estimate of the
rest frame equivalent width of the \nv\ line is $\simm 4.2\unit{\angstrom}$.

Given the presence of BAL troughs, it is difficult to infer very much
about the neutral hydrogen along the line--of--sight to \firstqso, as
the \lya\ emission line is all but completely absorbed.
Similarly, it is not possible to measure the \cite{Gunn_Peterson:1965}
optical depth to \firstqso: assuming the \nv\ BAL has a similar
velocity width to the \civ\ and \siiv\ BALs, the absorption blueward
of \lya\ is most likely dominated by the wing of the \nv\ BAL, and not by
intervening neutral hydrogen along the line--of--sight.

The case of \firstqso\ is rather similar to that of the BAL quasar
SDSS~J1044$-$0125 \citep{Fan_etal:2000}.  The initial redshift
measurement of $z=5.80$ by \cite{Fan_etal:2000} was revised to
$z=5.74$ by \cite{Goodrich_etal:2001} on the basis of NIR
spectroscopy, after which \cite{Jiang_etal:2007} used the \ciii\ line
to estimate $z=5.78$.  Both \firstqso\ and SDSS~J1044$-$0125 
demonstrate the importance of NIR spectra to obtaining reliable
redshifts of $z\simeq6$ quasars (and especially BALs).


\section{Conclusions and future prospects}
\label{section:conc}

Two new high-redshift quasars have now been discovered in the UKIDSS
LAS: \firstqso\ at $z = \zfirstqso$ (previously given as $z =
\zfirstqsoold$; \citealt{Venemans_etal:2007});
and \secondqso\ at $z =
\zsecondqso$.  Two previously known high-redshift quasars,
\firstsdssqso\ at $z = 5.82$ \citep{Fan_etal:2001} and \secondsdssqso\
at $z = 5.93$ \citep{Fan_etal:2004}, have also been recovered in the
UKIDSS DR3 LAS area.  The two new quasars are well within the
redshift range covered by optical surveys, but they 
were both too faint in SDSS to be included 
in the high-redshift quasar sample defined by 
\citet{Fan_etal:2003}.  
While the revised lower redshift of \firstqso\
puts its \lya\ emission near the peak of the SDSS \zprime\ band response, 
the line is almost completely absorbed by a \nv\ BAL, which results in a
reduction of the \zprime-band flux.  \secondqso\ has a very weak \lya\
line, with a rest-frame equivalent width of just $\simm20
\unit{\angstrom}$. The small \lya\ equivalent width results in the
\zprime-band \ston\ falling marginally below the limit of the SDSS
quasar sample defined by \cite{Fan_etal:2003}.  Nonetheless, the
detection of four $z \simeq 6$ quasars brighter than $Y_\vega = 19.88$
in the $870\unit{\sqdeg}$ of UKIDSS DR3 LAS is consistent with
expectations from the luminosity function of \cite{Jiang_etal:2008},
and so these new detections do not imply the existence of a previously
unidentified quasar population with weak emission features.

These results demonstrate that the UKIDSS data are of sufficient quality
to recover $z \ga 6$ quasars down to $Y \simeq 19.5$ with reasonable
completeness. Quantification of the completeness will be explored in
more detail when the sample is larger.  At the time of writing the
UKIDSS Fifth Data Release (DR5; \citealt{Warren_etal:2009}) has
already taken place, increasing the LAS area covered in the $Y$ and $J$
bands to $\simm1360\unit{\sqdeg}$.
DR5 is of sufficient size that finding no quasars of
redshift $z > 6.4$ would be somewhat inconsistent with expectations.


\begin{acknowledgements}

Many thanks to the staffs of UKIRT, the Cambridge Astronomical
Survey Unit, and the Wide Field Astronomy Unit, Edinburgh, for their
work in implementing UKIDSS.  Thanks to Marie Lemoine--Buserolle,
Kathy Roth and Claudia Winge for help in setting up the Gemini
observations.  Paul Dobbie kindly provided the synthetic spectrum of
GD50 used for flux-calibrating the optical spectrum of \firstqso.

Based on observations obtained at the Gemini Observatory (acquired through 
the Gemini Science Archive), which is operated by the Association of 
Universities for Research in Astronomy, Inc., under a cooperative agreement 
with the NSF on behalf of the Gemini partnership: 
the National Science Foundation (United States), 
the Science and Technology Facilities Council (United Kingdom), 
the National Research Council (Canada), CONICYT (Chile), 
the Australian Research Council (Australia), 
CNPq (Brazil) and SECYT (Argentina).

The Liverpool Telescope is operated on the island of La Palma by
Liverpool John Moores University in the Spanish Observatorio del Roque
de los Muchachos of the Instituto de Astrofisica de Canarias with
financial support from the UK Science and Technology Facilities
Council.

MP acknowledges support from the University of London's Perren Fund.
PCH, RGM and BV acknowledge support from the STFC-funded Galaxy
Formation and Evolution programme at the Institute of Astronomy.

The referee, Michael Strauss, made a number of valuable suggestions 
which have significantly improved this paper.

\end{acknowledgements}


\bibliographystyle{aa}
\bibliography{references}


\end{document}